# Development of a Bankruptcy Prediction Model for the Banking Sector in Mozambique Using Linear Discriminant Analysis

Reis Castigo Intupo[1]

[1] Faculty of Economics, Eduardo Mondlane University, Maputo, Mozambique

Correspondence: Reis Castigo Intupo, Faculty of Economics, Eduardo Mondlane University, Maputo, Mozambique.

**Abstract**

In Mozambique there is no evidence of a bankruptcy prediction model developed in the national economic context, yet, back in 2016, the national banking sector suffered a financial shock that resulted in Mozambique's Central Bank intervention in two banks (Moza Banco, S.A. and Nosso Banco, S.A.). This was a result of the deterioration of their financial and prudential indicators, although Mozambique had been adhering to the Basel Accords since 1994. The Basel Accords provides recommendations on banking sector supervision worldwide with the aim to enhance financial system stability. While it doesn't predict bankruptcy, the prediction model can be used as an auxiliary tool to manage that risk, but this has to be built in the national economic context. This paper develops for Mozambique's banking sector a bankruptcy prediction model in the Mozambican context through the linear discriminant analyses method, following two assumptions: (i) composition of the sample and (ii) robustness of the financial prediction indicators (the capital structure, profitability asset concentration and asset quality) from 2012 to 2020. The developed model attained an accuracy level of 84% one year before Central Bank intervention (2015) with the entire population of 19 banks of the sector, which makes it recommendable as a risk management tool for this sector.

## 1. Introduction

As per Mozambique's Central Bank report (2015 to 2020), the Mozambican banking sector was composed of 16 commercial banks at the end of 2020, 17 in 2019, 18 in 2018, 19 banks from 2015 to 2017 and 18 banks from 2012 to 2014. Their main role is to provide financial services to the economy, and according to the World Bank (2020), statistics data shows that domestic credit to the private sector represented 24% of Mozambique's GDP in 2020, 32.6% in 2015 and 22.9% in 2012. Comparing this funding to the economy with some African neighbors like Zimbabwe (15.2% in 2020, 18.3% in 2015 and 20.1% in 2012), Tanzania (12.4% in 2020, 14.6% in 2015 and 12.8% in 2012) and Zambia (5.4% in 2020, 19.8% in 2015 and 15.9% in 2012), it shows undoubtedly the importance of this sector to the Mozambican economy. While these countries also have in common the fact of having already experienced bank failure in their economies, none of them besides Zimbabwe has evidence of a working model developed internally to predict bankruptcy in its banking sector, despite early detection of bank failure concerning bank regulators worldwide.

Over the past decade, Mozambique officially reported that the national banking sector suffered financial shocks that in 2016 resulted in the Central Bank of Mozambique intervening in two banks (Moza Banco, S.A. and Nosso Banco, S.A). As mentioned by Associação Moçambicana de Bancos [AMB] (2016), this was due to the deterioration of their financial and prudential indicators, and unsustainable economic and financial structure, including serious liquidity and management problems. Moza Banco S.A was eventually recapitalized by the Bank of Mozambique, while Nosso Banco's was liquidated.

Bankruptcy is extremely costly, not just for the companies' owners, but also for all stakeholders, as its social-economic impact is extensive. That implies that less expensive alternatives to predict bankruptcy are very important. This is why many research efforts have been made to develop and improve prediction models. As defended by Citterio (2020: p.1), "prediction models can therefore be useful to predict if a business will suffer a financial distress and to define the related determinants through mathematical and statistical methods."

In Mozambique, no studies on the prediction of banking bankruptcy models developed using a domestic economic context were found and there is even no confirmation of the usage of different models developed internationally by the banking sector, despite the progressive and sophisticated techniques adopted by the banking sector regulator to evaluate credit, market and operational risks. These techniques are also aligned with the Basel I Accord and Basel III Accord, which Mozambique adopted, respectively in 1994 and 2013. However, it does not predict bankruptcy at all, but rather what is called the Internal Capital Adequacy Assessment Process (ICAAP), which helps the regulator Central Bank with international principles that stabilize the financial system and; thus avoid contagion risk. Nonetheless, this is only aligned with a good international macro-prudential approach.

The first multivariate statistical model is linked to Altman in 1968, known as the Z-score model, which is even used today to predict bankruptcy. Before Altman, many studies were based on Univariate modelling. However, Altman (1968) found that Univariate modelling is not enough to predict bankruptcy. Then, using Multivariate Discriminant Analysis (MDA), he analysed the risk of bankruptcy by combining different financial ratios of a group of manufacturing companies divided into two equal subgroups, classified as bankrupt and non-bankrupt during the period from 1946 to 1966. From this process, Altman developed a bankruptcy prediction model that had a predictability accuracy of 95% one year before bankruptcy and 72% for two years before bankruptcy. Altman (2000), however, after 30 years of tests of the z-score model, changed some ratios as a way to adjust the model for private, industrial and unlisted companies. However, came to the same conclusion reached when he developed his first model in 1968, that the ratios of bankrupt companies differ significantly from those of non-bankrupt companies, whose tendencies deteriorate the closer they get to bankruptcy, particularly between two and three years prior and that in general, the ratios that measure profitability, liquidity and solvency are the most significant for assessing the risk of bankruptcy in companies. According to Fejér-Király (2015) and Peres, and Antão (2017), since Altman, many researchers have used the discriminant analysis, whenever integrating or substituting new ratios to their models. Fejér-Király (2015) and Citterio (2020), also state that many other models based on statistical theory were also developed, such as logit analysis and probit analysis.

It is important to notice that even during and after the global financial crisis, different models are still being used today worldwide in different industries as well as several specific studies that emerged to analyse bank bankruptcy. According to Betz, Oprică, Peltonen and Sarlin (2013) several studies, mainly focusing on US banks, have recently emerged to analyse bank bankruptcy during the global financial crisis. The same authors also mentioned that multiple discriminant analysis methodology and artificial neural network in early-warning exercises find a high degree of predictability of US bank failures during the global financial crisis. Similar research for predicting US banks bankruptcy from 2008 to 2013, using logit and Canonical Discriminant Analysis models, also showed an accuracy over 86% during the global financial crisis period (Affes & Hentati-Kaffe, 2016). Nevertheless, for better performance, the models have to be developed in its economic context. As documented by Peres and Antão (2017) some general limitations that are common to the different prediction models that affect its performance, namely, territorial sensitivity, which is based on the fact that different countries have different legal requirements, accounting and financial systems, tax and labour systems, credit access policies, macro and microeconomic policies, and cultural issues. Sectorial sensitivity refers to the intrinsic financial characteristics among different sectors. Time sensitivity, meaning that the business reality from the period of the designed model might be different to the upcoming decade and therefore unlikely to yield the same performance on different periods. These limitations can also be found in a summary in a paper by Moody, when mentioning the heterogeneity in domestic legal, operating, financial reporting, and banking regulations between different countries (Wang, Dwyer, & Zhao, 2014).

The performance of the bankruptcy model also varies according to the approaches, creativity and skills of the researcher, including the financial ratios used for their validation, and the type of activity being analysed. Further to this, with the existence of several financial indicators available to analyse the risk of bankruptcy, the application of the statistical methodology suggests that it is possible to develop new models. In addition, there seems to be no consensus about which are the best models but rather that, no bankruptcy model specifies the date when the bankruptcy will occur, but instead, they alert to the risk, through symptoms gleaned from financial reports. Thus, given the great importance and interest to the bank's regulators, creditors and depositors on this matter, the purpose of this paper is to develop an early warning bankruptcy model that can detect such risk 12 months in advance, through discriminant analysis based on a set of financial ratios in the banking sector of Mozambique. Additionally, the intent of this article is also to promote future empirical research for preventing bank failures and financial crises, especially in Mozambique.

## 2. Method

*2.1 Description of the Sample and Sample Size*

The methodology for the selection of the sample is non-probabilistic for convenience. The model used 14 commercial banks in the sample, composed of 2 bankrupt and 12 non-bankrupt, through analysis of financial statements from 2012 to 2015. Additionally, with a financial statement from 2016 to 2020, 5 more banks out-of-sample were used to test the performance accuracy of the model, totalling 19 commercial banks, which is representative of the whole population of the Mozambique banking sector. This is in line with arguments presented by Peres and Antão (2017) that, with financial indicators normally distributed, the company under analysis is comparable to the one originally used to estimate the model.

The sample is composed of Moza Banco, Nosso Banco (both formally reported as bankrupted in 2016), Banco Internacional de Moçambique (BIM), Banco Comercial e de Investimentos (BCI), Standard Bank, Absa Bank Moçambique, Banco Terra, First National Bank de Moçambique, African Banking Cooperation Ecobank Moçambique, First Capital Bank, Société Générale Moçambique, Banco Nacional e de Investimentos and United Bank for Africa.

The banks out-of-sample also operating in Mozambique that have been taken into consideration are namely, Nedbank Moçambique, Capital Bank (Mozambique), Mybucks Banking Corporations, Socremo Banco Microcrédito and Banco Letshego, with financial statements from 2012 to 2020. They are not included as part of the sample due to a lack of data on either the first 2 or 3 years (Mybucks Banking Corporations, Banco Mais, Banco Letshego) or because their data were creating outliers (Nedbank Moçambique and Socremo Banco).

*2.2 The Modelling Approach*

2.2.1 Discriminant Function

According to Ayinla and Adekunle (2015: p. 13), discriminant analysis (DA), is a classic classification method, originally developed in 1936 by R.A Fisher, used to produce models whose accuracy approach often exceeds more complex and modern methods. These authors also mentioned that DA is very similar to logit regression with the only difference between them, being that "logit regression does not have as many assumptions and restrictions as DA". These assumptions and restrictions can be related to what Fejér-Király (2015) and Citterio (2020), consider the basic difference between the DA and the logit regression developed by Ohlson in 1980, which the logit regression does not take into consideration what DA proposes: normal distribution of the variables, and the variance and covariance matrix must be the same in the case of bankrupt and non-bankrupt firms. Nonetheless, Peres and Antão (2017) stated that small deviations or violations on normal multivariate distribution, and the homogeneity of variance and covariance matrices among the groups generally have no serious implications. Moreover, Ayinla and Adekunle (2015: p. 13) state that "when discriminant analysis' assumptions are met, it is more powerful than logit regression."

The model of this study is based on linear discriminant analysis (LDA), using financial categories of profitability, leverage, asset quality and asset concentration. As per Ayinla and Adekunle (2015: p.12), discriminant analysis "…is sometimes preferable to logit regression especially when the sample size is very small and the assumptions are met." Additionally, as per Poulsen and French (2018) with discriminant assumption, it is not mandatory to have an equal size between groups in the sample.

The designed statistical model of bankruptcy prediction for the banking sector in Mozambique was specified as a regression model as follows:

$$Z = a + b_1 X_1 + b_2 X_2 + \ldots + b_n X_n \qquad (1)$$

Where: (Z) discriminating index or bankruptcy factor (which corresponds to a binary/dummy variable); (a) constant of discriminant function; ($b_i$) the discriminant coefficient or weight for independent variable; ($X_i$) the independent (or predictor) variable; and (n) the number of independent variables.

As "Z" is a binary discriminating index, the given estimation value is 0 (zero) for bankrupt banks (risk of bankruptcy) and 1 (one) for non-bankrupt banks (or bankruptcy unlikely to happen), which only gives 2 (two) classification zones.

2.2.2 Selection of financial ratios (independent variables)

The selection of variables was based on the adjustments of the ratios of the four categories of financial ratios (capital structure ratio, profitability ratio, asset concentration and asset quality). The choice of these categories, specifically the capital structure and profitability is reinforced by the fact that they mirror the company's financial autonomy and the company's ability to generate returns for both lenders and shareholders, respectively. Furthermore, as per Singhal, Goyal, Sharma, Kumari, and Nagar (2022), capital ratios are used to quantify a bank's capitalization; and ROA and ROE are widely used to assess bank's profitability. Regarding asset concentration and asset quality, these were chosen because they reflect the risk when maintaining certain levels of volatile assets in the banking sector. These ratios are frequently used by Moody to assess banks risk default (Wang et al., 2014).

Table 1. Case processing summary

| Healthy bank status | | Cases | | | | | |
|---|---|---|---|---|---|---|---|
| | | Valid | | Missing | | Total | |
| | | N | Percent | N | Percent | N | Percent |
| EAA | Bankrupt | 2 | 100% | 0 | 0% | 2 | 100% |
| | Non bankrupt | 12 | 100% | 0 | 0% | 12 | 100% |
| ROAA | Bankrupt | 2 | 100% | 0 | 0% | 2 | 100% |
| | Non bankrupt | 12 | 100% | 0 | 0% | 12 | 100% |
| ROAE | Bankrupt | 2 | 100% | 0 | 0% | 2 | 100% |
| | Non bankrupt | 12 | 100% | 0 | 0% | 12 | 100% |
| NII | Bankrupt | 2 | 100% | 0 | 0% | 2 | 100% |
| | Non bankrupt | 12 | 100% | 0 | 0% | 12 | 100% |
| LAAA | Bankrupt | 2 | 100% | 0 | 0% | 2 | 100% |
| | Non bankrupt | 12 | 100% | 0 | 0% | 12 | 100% |
| BDTL&A | Bankrupt | 2 | 100% | 0 | 0% | 2 | 100% |
| | Non bankrupt | 12 | 100% | 0 | 0% | 12 | 100% |

Table 1 depicts the six independent variables taken into consideration in the model for both groups, namely, Capital structure ratio: equity to average total assets (EAA); Profitability: return on average equity (ROAE); Profitability: return on average total assets (ROAA); Profitability: net interest income by average total assets (NII); Asset concentration ratio: loans and advances to average total assets (L&AAA); and Asset Quality: bad debts to total loans and advances (BDTL&A). Net interest income and Asset concentration data were collected from Klynveld Peat Marwick Goerdeler [KPMG] banking survey reports over the period 2012 – 2020. Additional data were extracted from the respective bank's website for the period under study.

The number of variables in the models will be according to Altman (1968) assumptions, which consider the maximum number as equal to the sample size minus 1.

*2.3 Diagnostic Tests and Model Validation*

The diagnostic test was made with the SPSS statistic 29 package, and discriminant analysis followed the testing of its restrictive assumptions. According to Citterio (2020), there are three main restrictive assumptions, namely (i) normal distribution of explanatory variables; (ii) equal variance-covariance matrices across the groups; (iii) absence of multicollinearity.

2.3.1 Normal Distribution of Explanatory Variables

Table 2. Average ratios of independent variables

| Bank | EAA | ROAE | ROAA | NII | L&AAA | BDTL&A |
|---|---|---|---|---|---|---|
| Moza Banco, S.A | 0.13386 | 0.03166 | 0.00044 | 0.04560 | 0.74644 | 0.02260 |
| Nosso Banco, S.A | 0.01869 | 0.12351 | -0.01210 | 0.03388 | 0.68139 | 0.03105 |
| Banco Internacional de Mocambique | 0.17837 | 0.23572 | 0.03902 | 0.06100 | 0.63682 | 0.02600 |
| Banco Comercial e de Investimentos | 0.06885 | 0.37131 | 0.01794 | 0.04000 | 0.63848 | 0.01350 |
| Standard Bank, S.A | 0.16606 | 0.22077 | 0.03386 | 0.05525 | 0.59744 | 0.02100 |
| Barclays Bank Moçambique, S.A | 0.15588 | -0.13747 | -0.01820 | 0.05425 | 0.48868 | 0.08075 |
| Banco Terra, S.A | 0.38839 | -0.42163 | -0.10999 | 0.06375 | 0.74355 | 0.23375 |
| FNB Moçambique, S.A | 0.15968 | 0.07917 | 0.00917 | 0.06950 | 0.65537 | 0.04000 |
| African Banking Cooperation, S.A | 0.10499 | -0.00501 | 0.00046 | 0.05450 | 0.62708 | 0.10225 |
| Ecobank Moçambique, S.A | 0.21022 | -0.28309 | -0.05792 | 0.12833 | 0.37958 | 0.07000 |
| First Capital Bank | 0.31912 | -0.28404 | -0.07633 | 0.03833 | 0.49022 | 0.24333 |
| The Mauritius Bank | 0.20648 | 0.00387 | -0.00688 | 0.03125 | 0.33394 | 0.13600 |
| Banco Nacional e de Invest. | 0.75149 | 0.05102 | 0.02961 | 0.08118 | 0.14627 | 0.00145 |
| United Bank for Africa | 0.09803 | -0.23551 | -0.15304 | 0.07700 | 0.59878 | 0.20500 |
| Average | 0.21144 | -0.01784 | -0.02171 | 0.05956 | 0.55457 | 0.08762 |

Table 2 shows the average values from 2012 to 2015 of each independent variable for the banks considered in the sample. This table also represents the data to be transformed to z-score normalization. While according to Ayinla and Adekunle (2015), and Citterio (2020), the normality assumption usually may not compromise the model.

Table 3. Z-score normalization of independent variables

| Bank | Z-score EAA | Z-score ROAE | Z-score ROAA | Z-score NII | Z-score L&AAA | Z-score BDTL&A |
|---|---|---|---|---|---|---|
| Moza Banco, S.A | -0.42518 | 0.37956 | 0.21827 | -0.54198 | 1.12617 | -0.76404 |
| Nosso Banco, S.A | -1.05727 | 0.16980 | 0.62162 | -1.02120 | 0.74392 | -0.66949 |
| Banco Internacional de Moçambique | -0.18334 | 1.06128 | 1.11266 | 0.05705 | 0.48517 | -0.72858 |
| Banco Comercial e de Investimentos | -0.78245 | 0.69420 | 1.70453 | -0.78159 | 0.49105 | -0.87042 |
| Standard Bank, S.A | -0.24930 | 0.97388 | 1.04689 | -0.18256 | 0.24993 | -0.78768 |
| Barclays Bank Moçambique, S.A | -0.30426 | 0.06493 | -0.52266 | -0.22250 | -0.38519 | -0.07851 |
| Banco Terra, S.A | 0.97090 | -1.54322 | -1.77217 | 0.17686 | 1.11441 | 1.72986 |
| FNB Moçambique, S.A | -0.28228 | 0.53688 | 0.42433 | 0.41647 | 0.59102 | -0.56311 |
| African Banking Cooperation, S.A | -0.58458 | 0.37956 | 0.05606 | -0.18256 | 0.42636 | 0.16969 |
| Ecobank Moçambique, S.A | -0.00746 | -0.63427 | -1.16276 | 2.73270 | -1.02620 | -0.20853 |
| First Capital Bank | 0.59165 | -0.94891 | -1.16715 | -0.86146 | -0.37931 | 1.83623 |
| The Mauritius Bank | -0.02944 | 0.25720 | 0.09551 | -1.14100 | -1.29671 | 0.57155 |
| Banco Nacional e de Invest. | 2.96608 | 0.90396 | 0.30157 | 0.85575 | -2.40230 | -1.02407 |
| United Bank for Africa | -0.62305 | -2.29486 | -0.95670 | 0.69601 | 0.26170 | 1.38709 |

Table 3 depicts the transformation of the data into z-score normalization data in SPSS which each predictor within the sample has a mean of 0 (zero) and standard deviation of 1 (one) to avoid outlier issues. These are the datasets for processing the analysis.

2.3.2 Absence (or Low Level) of Multicollinearity

This was tested by a correlation matrix within the groups. According to Ayinla and Adekunle (2015), if the value of correlation among two or more explanatory variables exceeds 0.8 (80%) then the variables are highly collinearly related, and this can compromise the reliability of prediction of group membership.

2.3.3 Significance of Discriminant Function

This was tested by Wilks' Lambda. As per Stella (2019), Wilks' Lambda table is used to indicate the significance of the discriminant function. Its value varies between 0 (zero) and 1 (one), and the closer to 0 (zero) the more preferred is the discriminant function, as the lower value implies greater importance the independent variables have to the discriminant function, meaning a higher significance of the discriminant function. On the other hand, a small Wilks' Lambda value indicates that the group means appear to differ. For this analysis the null hypothesis ($H_0$) implies the group's means are equal if (Sig) ≥ 0.05, otherwise the alternative hypothesis ($H_1$) is accepted, inferring that the group's means are not equal.

2.3.4 Homogeneity (or Homoscedasticity)

Equal variance-covariance matrices across the groups were tested through Box's M statistic, in which the null hypothesis ($H_0$) implies the entire population variance within the groups is equal if alpha (Sig) ≥ 0.05, otherwise the alternative hypothesis ($H_1$), is accepted, inferring that the variance between the groups is not equal. As defended by Ayinla and Adekunle (2015), and Citterio (2020), the covariance matrix within each group should be equal, although this is not mandatory.

2.3.5 Parameter of Canonical Correlation

Canonical correlation provides the variation between the discriminant functions ($R^2$) and explanatory variables given by the eigenvalues table (Stella, 2019). It was tested by squaring the canonical correlation. Results closer to 1 depict a more discriminant function.

2.3.6 Canonical Discriminant Function and Discriminant Coefficient

The bankruptcy model is a regression equation given by the unstandardized coefficients from the canonical discriminant function (Stella, 2019).

The standardized discriminant coefficient indicates the discriminant weight of each independent variable within the canonical discriminant function (Stella, 2019). Variables with high discriminant weight are usually more discriminatory. Additionally, as mentioned by Ayinla and Adekunle (2015) which independent variable discriminates more is one of the basic questions for evaluation criteria for discriminant analysis.

2.3.7 Classification Matrix and Cross Validation

The classification result output is given as the simple summary of cases classified correctly and incorrectly. This is given by a classification table, also called a confusion table (Stella, 2019). It is expected a correct classification close to 100%.

2.3.8 Classification Zones and Overall Robustness of the Model

While the cut-off point only provides 2 (two) classification zones, then the third zone (grey zone) is estimated to reduce the classification error, mainly of type I error. Thus, given the group centroid of both groups, the cut-off point is given as:

$$\text{Cut-off point} = (Y_0 N_0 + Y_1 N_1) / (N_0 + N_1) \qquad (2)$$

Where: $Y_0$ and $Y_1$ represent the discriminant mean (group centroid) for the bankrupt and non-bankrupt group, respectively; $N_0$ and $N_1$ represent the group size for bankrupt and non-bankrupt, respectively.

Three (3) zones represent the classification zones, which are calculated based on the property of normal distribution, through sum and subtracts the respective centroid group with the respective standard deviation of each group. The three zones are namely, bankrupt (bankruptcy likely to occur), non-bankrupt (bankruptcy unlikely to occur) and grey zone (which means the potential for bankruptcy is uncertain).

The overall classificatory ability is the hit ratio percentage given as a hit number (correctly classified) by the total number of observations each year.

## 3. Results and Discussion

*3.1 Statistics and Data Analysis*

3.1.1 Testing normality of the predictors variables

Table 4. Group statistics

|   |   |   |   |   | Valid N (listwise) | |
|---|---|---|---|---|---|---|
|   |   |   | Mean | Std. Deviation | Unweighted | Weighted |
| Total | Z-score: | EAA | 0.0000000 | 1.00000000 | 14 | 14.000 |
|   | Z-score: | ROAA | 0.0000000 | 1.00000000 | 14 | 14.000 |
|   | Z-score: | ROAE | 0.0000000 | 1.00000000 | 14 | 14.000 |
|   | Z-score: | NII | 0.0000000 | 1.00000000 | 14 | 14.000 |
|   | Z-score: | LAAA | 0.0000000 | 1.00000000 | 14 | 14.000 |
|   | Z-score: | BDTL&A | 0.0000000 | 1.00000000 | 14 | 14.000 |

As shown in Table 4, all independent variables for both groups of banks (bankrupt and non-bankrupt) are z-score normalized, which makes the normal distribution assumption met with a mean of 0 (zero) and standard deviation of 1 (one).

3.1.2 Multicollinearity Test

Table 5. Correlation matrix

| Correlation |   | Z-score: EAA | Z-score: ROAA | Z-score: ROAE | Z-score: NII | Z-score: LAAA | Z-score: BDTL&A |
|---|---|---|---|---|---|---|---|
|   | Z-score: EAA | 1.000 | 0.066 | -0.225 | 0.210 | -0.606 | -0.022 |
|   | Z-score: ROAA | 0.066 | 1.000 | 0.834 | -0.222 | -0.195 | -0.859 |
|   | Z-score: ROAE | -0.225 | 0.834 | 1.000 | -0.345 | 0.040 | -0.799 |
|   | Z-score: NII | 0.210 | -0.222 | -0.345 | 1.000 | -0.221 | -0.178 |
|   | Z-score: LAAA | -0.606 | -0.195 | 0.040 | -0.221 | 1.000 | 0.222 |
|   | Z-score: BDTL&A | -0.022 | -0.859 | -0.799 | -0.178 | 0.222 | 1.000 |

The result of Table 5 shows that there is a significant positive correlation between ROAA and ROAE, which means that as one variable increases so does the other. This suggests removing one of these two predictors from the matrices because of redundant performance between them. On the other hand, ROAA and BD revealed a significant negative correlation between them, which in fact makes sense as a high level of bad debts (BD) reduces the profitability. This negative impact of BD to profitability's indicators can also be seen with ROAE whose correlation is close to 80%. These findings of negative correlation are in line with a study conducted by Al-Sharkas and Al-Sharkas (2022) and Naili and Lahrich (2022).

Even though with low collinearity, Table 5 also shows a negative relationship between bad debt (BD) and capital structure (EAA), which is expected as non-performing loans (NPL) may shrink the bank's capitalization. These results corroborate with the finding from Naili and Lahrich (2022) argument, that the banks usually raise their capital to face risk exposure. On the other hand, return on equity (ROEA) and capitalization (EAA) presented an unexpected negative relationship. This unexpected result is supported by assuming that earnings can be used as funding sources; thus increasing profitability is expected to better capital ratios. For instance, from 2000 to 2020 Singhal, et al. (2022) found considerable positive effects while analysing the influence of profitability (ROA and ROE) on capitalization in some BRICS banking sector, namely, banks from Brazil, Russia, and India. While for China and South Africa, the authors found a negative effect between these ratios. Naili and Lahrich (2022) found from research in MENA non-GCC countries (Morocco, Tunisia, Egypt, Jordan and Turkey) that negative interrelationships can be explained by the fact that high agency costs might result in an increased level of non-performing loans, and because the shareholders may demand high return; therefore, both affecting profitability.

### 3.1.3 Significance of Discriminant Function

Table 6. Discriminant function test

| Test of Function(s) | Wilks' Lambda | Chi-square | Df | Sig. |
|---|---|---|---|---|
| 1 | 0.242 | 12.778 | 6 | 0.047 |

Table 6 shows the discriminant function test with 6 degrees of freedom and a p-value of 0.047. Since the p-value is less than 0.05, and Wilks lambda value of 0.242 is close to 0, the null hypothesis is rejected ($H_0$) and it is then inferred that the discriminant function is significant and can be used for further interpretation of the result. Thus, this assumption is met for discriminant analysis.

### 3.1.4 Homoscedasticity

Table 7. Homogeneity test

| Box's M | | 4.416 |
|---|---|---|
| F | Approx. | 3.722 |
| | df1 | 1 |
| | df2 | 26.596 |
| | Sig. | 0.064 |

Tests null hypothesis of equal population covariance matrices of canonical discriminant functions.

Box's M statistic test was used as shown in Table 7, resulting in a p-value of 0.064 which is higher than 0.05, thus the null hypothesis ($H_0$) is accepted, reflecting equal variance-covariance matrices between the groups (discriminant function), and the variance of each cluster is homogenous. Therefore, this assumption is met for discriminant analysis.

### 3.1.5 Canonical Correlation

Table 8. Canonical correlation test

| Function | Eigenvalue | % of Variance | Cumulative % | Canonical Correlation |
|---|---|---|---|---|
| 1 | 3.136[a] | 100 | 100 | 0.871 |

a. First 1 canonical discriminant functions were used in the analysis.

As shown in Table 8, the square of canonical correlation $(0.871)^2$ is equal to 75.8%, which reflects the percentage of the discriminant function variation explained by the predictors. As the value is close to 100%, it gives the condition to move forward with the analysis.

### 3.1.6 Canonical Discriminant Function and Discriminant Coefficients

Table 9. Bankruptcy model and discriminant weight

| Canonical Discriminant Function Coefficients | Function 1 | Standardized Canonical Discriminant Function Coefficients | Function 1 |
|---|---|---|---|
| Z-score: EAA | -0.040 | | -0.040 |
| Z-score: ROAA | 2.151 | | 2.224 |
| Z-score: ROAE | 2.548 | | 2.609 |
| Z-score: NII | 2.377 | | 2.335 |
| Z-score: LAAA | -0.487 | | -0.466 |
| Z-score: BDTL&A | 4.734 | | 4.694 |

Unstandardized coefficients.

Table 9 depicts the developed model to assess the health bank status. This discriminant function maximizes the difference between both groups which provides the discriminant model as follows:

$$Z = -0.040 EAA + 2.151 ROAA + 2.548 ROAE + 2.377 NII - 0.487 LAAA + 4.734 BDTL\&A \qquad (3)$$

The same table also provides the discriminatory weight of each predictor in the discriminant function in terms of which variable discriminates more between bankrupt and non-bankrupt group. The sign indicates the direction of the relationship in the respective group. Thus, it can be seen that the BDTL&A discriminate more than other variables, followed by the ROAE, NII, ROAA, L&AAA and finally the EAA. In a nutshell, the sequence of importance of the categories is as follows; asset quality, profitability, liquidity and leverage. Having EAA as one of the least discriminatory ratios in terms of weight and accompanied by a negative sign, it can then be inferred that during the period of analysis, the banking sector performed with low levels of financial autonomy and consequently high levels of financial leverage, resorting to debt for their financing, leading to more interest payments, which in turn has implications on profitability.

Gumbo and Zoromedza (2016) using the logit model with non-performing loans ratio, return on equity, and other different variables to predict failure for Zimbabwean banks between 2009 and 2013 also identified the non-performing loans with positive coefficient as the major discriminator of bank probability of failure when comparing to ROE. Furthermore, if comparing the importance of the non-performing loan to total asset ratio with the loan and advances to total asset and return on asset, comparative research to predict bank failure in Nigeria, using Cox Proportional Hazards Model and data covering a period from 2003 to 2011 by Babajide, Olokoyo and Adegboye (2015) also highlighted the same finding.

Table 10. Group membership classification function

|  |  | Health bank status | |
|---|---|---|---|
|  |  | Bankrupt | Non-bankrupt |
| Z-score: | EAA | 0.162 | -0.027 |
| Z-score: | ROAA | -8.639 | 1.440 |
| Z-score: | ROAE | -10.232 | 1.705 |
| Z-score: | NII | -9.548 | 1.591 |
| Z-score: | LAAA | 1.957 | -0.326 |
| Z-score: | BDTL&A | -19.010 | 3.168 |
| (Constant) |  | -10.010 | -0.378 |
| Fisher's linear discriminant functions | | | |

Table 10 shows the classification function coefficient that can be used specifically for each group membership: On the bankrupt discriminant function, it can be noted the expected negative relationship of ROAA, ROAE and NII ratios, as in fact, the lower their values are, the more likely the risk of bankruptcy is likely to happen. On the other hand, for the non-bankrupted banks, the same ratios presented an expected positive relation with the dependent variable. These ratios are related to the profitability category, which is in accordance with Dwyer, Kocagil and Stein (2004), Wang et al. (2014) and Gumbo and Zoromedza (2016) that high profitability reduces the probability of bankruptcy.

As per Dwyer, et al. (2004), the probability of bankruptcy diminishes as the ROA increases. Obadire, Moyo and Munzhelele (2022) argue that a positive and higher net interest margin shows good efficiency of the bank. Regarding the LAAA, as per Wang et al. (2014), ratios that measure the extent to which the bank has liquid assets relative to assets and liabilities, such as loans, might impact the bank's healthy, as high liquidity reduces bankruptcy probability. However, loans and advances are considered volatile assets, and as per Dwyer, Guo and Hood (2006), Iannotta, Nocera and Sironi (2007), Babajide et al. (2015), and Gumbo and Zoromedza (2016) the poor quality of loans granted can negatively impact the bank's profitability, since it is partly and directly related to the quality of their assets. Therefore, the recommendation is to pay attention to the concentration of volatile assets since these can lead to bank failure. Also, according to KPMG (2016), loans and advances to customers represented an increase in assets of around 58% and 59% in 2015 and 2014, respectively. They also reported that the credit quality in general deteriorated from 2014 to 2015, having increased from 3.9% to 4.01%, respectively. The average ratio of non-performing loans in small banks was around 7%, 2.99% above the percentage of the credit sector in general, which probably shows a weak effectiveness in the risk management process in the related banks. Such a situation can be seen in this table as bankrupt banks have relatively higher levels of volatile asset concentration (and consequently bad debts negatively impacted the bank's healthy) than the non-bankrupted banks.

Equity to average total assets ratio (EAA) is related to the bank's capital strength, which in fact implies a negative relationship with the probability of bankruptcy, nonetheless, it presented an unexpected performance in both groups during the period of 2012 to 2015. According to Santillán-Salgado (2015) and Obadire et al. (2022), an adequate capital buffer protects and cushions the bank from financial and economic stress like market risk, operational risk and credit risk, and so the risk of bankruptcy. This unexpected performance reflects the implications of the negative interrelationship of return on equity and non-performing loan to the capitalization of Mozambique's banking sector during the period under analysis.

3.1.7 Classification Matrix and Cross Validation

Table 11. Classification Results[a]

|  |  |  | Predicted Group Membership | |
|---|---|---|---|---|
| Health bank status |  |  | Bankrupt | Non-bankrupt |
| Original | Count | Bankrupt | 2 | 0 |
|  |  | Non-bankrupt | 0 | 12 |
|  | % | Bankrupt | 100.0 | 0.0 |
|  |  | Non-bankrupt | 0.0 | 100.0 |
| a. 100.0% of original grouped cases correctly classified. | | | | |

As shown in Table 11 the classification result is 100% composed of correctly classified group membership. Therefore, all 14 observations were correctly predicted by the model. With this result, the developed model has the condition to extend the assessment to the whole representative population of the Mozambique banking sector (19 banks) from 2014 to 2020 in order to have an overall classificatory ability of the model.

3.1.8 Classification Zones and Overall Robustness of the Model

Table 12. Group Centroid

| Health bank status | Function 1 |
|---|---|
| Bankrupt | -4.016 |
| Non bankrupt | 0.669 |
| Unstandardized canonical discriminant functions evaluated at group means. | |

Table 12 shows the means of the discriminant function scores by group. From this point, the cut-off point can then be calculated and used to estimate the intervals for three (3) classification zones.

Table 13. Cut-off point and classification zones

| Bank | Z-score | Group centroid | Cut-off point |
|---|---|---|---|
| Moza Banco, S.A | -4.0643 | -4.0160 | |
| Nosso Banco, S.A | -3.9678 | | |
| Banco Internacional de Mocambique | 1.5754 | | |
| Banco Comercial e de Investimentos | -0.3502 | | |
| Standard Bank, S.A | 0.4877 | | |
| Barclays Bank Moçambique, S.A | -1.8925 | | |
| Banco Terra, S.A | 0.1921 | | -0.000007 |
| FNB Moçambique, S.A | 0.2839 | 0.6690 | |
| African Banking Cooperation, S.A | 1.1443 | | |
| Ecobank Moçambique, S.A | 1.6832 | | |
| First Capital Bank | 1.7902 | | |
| The Mauritius Bank | 1.4226 | | |
| Banco Nacional e de Invest. | 0.9507 | | |
| United Bank for Africa | 0.7444 | | |
| *Classification zones* | Interval | | |
| Bankruptcy is likely to occur | Z < - 0.040 | | |
| Bankrupcy can not be predicted (Grey zone) | - 0.040 ≤ Z ≥ - 0.003 | | |
| Bankruptcy is not likely to occur | Z > - 0.003 | | |

Given the canonical discriminant function coefficients depicted in Table 13, the z-score is then calculated for each bank, which shows the group centroid that coincides with the group centroid from the SPSS output. Through this result the cut-off point is 0.0007%, giving just two classification zones. Rather than classify with only two zones, the third zone (grey zone) is estimated as shown in Table 13. As a result, banks whose discriminant index is between the values [-0.040; -0.003] are classified as being in the grey zone. Banks that are below and above the limit of the grey zone are classified respectively as banks with a high risk of bankruptcy (or bankrupted) (Z < -0.040) and with a lower risk of bankruptcy or non-bankrupted (Z> -0.003).

Table 14. Overall classification

| | 2020 | 2019 | 2018 | 2017 | 2016 | 2015 | 2014 | 2013 | 2012 |
|---|---|---|---|---|---|---|---|---|---|
| Bankrupt | 2 | 0 | 2 | 3 | 4 | 4 | 5 | 3 | 7 |
| Grey zone | 1 | 0 | 0 | 0 | 1 | 0 | 0 | 1 | 0 |
| Nonbankrupt | 14 | 17 | 16 | 15 | 12 | 15 | 11 | 12 | 7 |
| Hit numbers | 14 | 17 | 16 | 15 | 12 | 16 | 11 | 12 | 7 |
| Total | 17 | 17 | 18 | 18 | 17 | 19 | 16 | 16 | 14 |
| Grey zone (%) | 6% | 0% | 0% | 0% | 6% | 0% | 0% | 6% | 0% |
| Type I error (%) | 0% | 0% | 0% | 0% | 0% | 50% | 0% | 0% | 0% |
| Type II error (%) | 12% | 0% | 11% | 17% | 24% | 12% | 31% | 19% | 50% |
| *Accuracy (%)* | *82%* | *100%* | *89%* | *83%* | *71%* | *84%* | *69%* | *75%* | *50%* |
| Bankrupt | 3 | 0 | 2 | 3 | 5 | 4 | 5 | 4 | 7 |
| Nonbankrupt | 14 | 17 | 16 | 15 | 12 | 15 | 11 | 12 | 7 |
| Hit numbers | 14 | 17 | 16 | 15 | 12 | 16 | 11 | 12 | 7 |
| Type I error (%) | 0% | 0% | 0% | 0% | 0% | 50% | 0% | 0% | 0% |
| Type II error (%) | 18% | 0% | 11% | 17% | 29% | 12% | 31% | 25% | 50% |
| *Accuracy based on Cut-off point* | *82%* | *100%* | *89%* | *83%* | *71%* | *84%* | *69%* | *75%* | *50%* |

Table 14 gives the general classification which summarizes the percentage of each classification zone, type error percentages and accuracy percentage. In both classification methods, through three zones classification and cut-off point, the model accuracy remains equal throughout all periods, with a maximum accuracy level of 100% in 2019. However, the cut-off point method appears with more cases of type II error, through the years 2013, 2016 and 2020. This means the cut-off point method in the mentioned years, has more cases of misclassified non-bankrupted banks as bankrupted banks, despite the equality of type I error in both methods. Therefore, the three zones method classifies better than the cut-off point method, even if the accuracy is the same.

Pointing to 2015, the model was supposed to classify two banks as bankrupted (Moza Banco and Nosso Banco), but instead, Nosso Banco, was miss classified as a healthy bank, leading the ratio to ½, which means 50% of type I error in the unhealthy bank group. On the other hand, the model also misclassified two banks that were supposed to be healthy banks rather than unhealthy banks, leading to a 12% of type II error. Nevertheless, the accuracy of 84% in both methods can be considered a good performance.

Indeed, from 2015 to 2012 this model shows the same characteristics of decreasing its accuracy from the year of bankruptcy backward as which has been developed by Altman since 1968.

In general, as per the banking survey of KPMG (2014, 2015, 2016, 2019 and 2022), the amount of loans and advances have grown over the years (especially Standard Bank, Capital Bank, Moza Banco, Banco Nacional de Investimentos, and Banco Mais in 2015), representing more than 50% of the total assets. The loans and advance-to-deposit ratios have also grown to more than 70% over the years, reflecting on the increased demand for loans to finance the economy. However, the non-performing loans have also increased over the years with an average of 4%. Moza Banco, one that was considered bankrupted by Mozambique's Central Banks in 2016, appeared in the KPMG report (2016), as one of the banks that registered significant growth in total assets, particularly the loans and advances in 2015, growing 31.6% at a slower pace when compared to the growth of total assets (43.6%) relative to the year 2014. This led to a decrease in the level of concentration of assets by 6.19% points from 2014 to 2015, thus contributing to the improvement of its LAAA in 2015. However, this decrease was not enough to offset the area of the greatest risk of bankruptcy, culminating with the fact that increased its financial leverage between these two years, together with a negative impact of profitability ratios, resulting in the decrease of its discriminant index.

Regarding Nosso Banco, after being in a zone of bankruptcy likely to occur in 2012, it upgraded its health status over the years to a zone of bankruptcy unlikely to occur in 2015. According to the study from KPMG (2016), this bank was in the list of the banks that registered a remarkable growth in total assets in 2015 but did not appear in the list of banks that recorded significant increases in loans and advances. Therefore, the loans did not contribute to increasing its assets, despite that the credit portfolio captured by BDTL&A was increasing over the years. However, profitability ratios performed relatively well from 2013 onwards, when compared to the previous year. This scenario might have contributed to the classification of this bank as a zone of bankruptcy unlikely to occur in 2015. Due to these facts, the model did not capture the reason for this bank to be liquidated, which makes this result unexpected as it was liquidated based on the same reason as Moza Banco. This might be the reason for some financial analysts along with media comments and speculation as to the reason to liquidate Nosso Banco. Voz da America [VOA] (2016b) argued "What should be behind the decision to liquidate Nosso Banco, which in 2015 was part of the group of banks that recorded remarkable growth?" (see Appendix A for complete proofs).

More comments arose, and according to VOA (2016a) citing the Center for Public Integrity, at least 5 banks were in the same situation as liquidated Nosso Banco, namely Capital Bank, African Banking Cooperation, Ecobank Moçambique, Banco Mais Moçambique, and United Bank of Africa (UBA). With the exception of African Banking Cooperation, the model has classified these banks to a zone of bankruptcy likely to occur between 2015 and 2016 (see Appendix B for complete proofs).

It is hoped that this research will provide a basis for future research as building a bank bankruptcy prediction model is a continuous process of research as there are no models and predictors considered unique and universal. Moreover, the univariate analysis methods are decreasing in usage vs. the technique of conjugating ratios to compose early warning bankruptcy models with recent versions using as few predictors as possible, but with an accuracy level approaching 100%.

One of the limitations of this analysis is the small sample size. As advocated by Poulsen and French (2018), the smallest group in the sample needs to exceed the number of independent variables, although the low sample size may work, it's not encouraged. Nevertheless, this limitation was particularly managed with the normalization of the sample dataset in order to avoid bias error situation between group membership.

Another limitation is related to the collinearity issue, especially between ROAA and ROAE which suggests excluding one of these indicators for the next analyses or conducting the analysis using a different method like logit regression which doesn't rely on multicollinearity assumption. However, both limitations did not compromise the reliability of the prediction of group membership as the finding in this study shows that the estimation model correctly classified group membership.

## 4. Conclusion

Despite the limitations of the sample size and some collinearity issues, particularly between ROAA and ROAE the model highlighted the key aspect of bank performance, and the working model represented in this paper includes only six predictors with an accuracy of 84% in 2015, one year before the intervention of Mozambique's Central Bank in Moza Banco and Nosso Banco, and its accuracy maintained the robustness over the years to 2020. Bad debts which are related to non-performing loans, was the critical predictor in determining the risk of bankruptcy for Mozambique's banking sector, as it has had direct implications on bank assets as loan and advance represented more than 50% of the Mozambican bank's total assets, especially from 2012-2015. This can represent serious negative implications to the economy, given that the domestic credit to the private sector represents more than 25% of Mozambique's GDP. On the other hand, profitability was the second most critical predictor, especially ROAE which can reflect losses as a direct result of impairment losses on loans and advances. Further to this, the research provides evidence of a negative implication of return on equity (ROAE) and bad debts (BDTL&A) on bank capital, affecting the capacity of a bank to face diverse risks like operational risk, market risk and credit risk. These findings are consistent with other research, particularly in some African countries.

Some researchers often attempt to assess potential bankruptcy using models that have already been disseminated in a different economic context rather than developing one. Others compare different models to determine which one has better accuracy. Moreover, some authors don't give a descriptive analysis of how to develop a model but only assumptions and limitations. Nonetheless, this paper provides a good starting point in predicting bankruptcy along with descriptive and summary statistics which has found an ideal combination of predictors that can prevent serious negative implications, particularly to the Mozambican banking sector and consequently to its economy as a whole. Additionally, it seems to be the first empirical study in predicting bankruptcy in Mozambique's banking sector, thus this study contributes to the literature with a new approach to risk management and serves as a comparative source for future research. However, as this paper is focused on quantitative indicators, future research attempts to combine quantitative and qualitative indicators as well as include different macroeconomic variables is encouraged.

This model serves as an auxiliary tool and provides very valuable information to the decision-makers and other stakeholders in risk management, which make it recommendable to be used in the banking sector of Mozambique.

**Acknowledgments**

I greatly appreciate the valuable contribution and comments from:

Professor Fernando Lichucha: University of Eduardo Mondlane. For proofreading the first draft of article and generous comments;

Professor Matias Farahane: University of Eduardo Mondlane. For proofreadind the first draft of article and generous comments;

David Wheeler: University of Lisbon. For proofreading the article, providing language help and generous comments;

**Authors contributions**

Reis Castigo Intupo, is only author responsible for study design, revising, data collection, drafted the manuscript, revising, and all process related to this manuscript. No any special agreements concerning authorship.

**Funding**

Not applicable

**Competing interests**

The author declare that he has no known competing financial interests or personal relationships that could have appeared to influence the work reported in this paper.

**Informed consent**

Obtained.

**Ethics approval**

The Publication Ethics Committee of the Redfame Publishing.

The journal's policies adhere to the Core Practices established by the Committee on Publication Ethics (COPE).

**Provenance and peer review**

Not commissioned; externally double-blind peer reviewed.

**Data availability statement**

The data that support the findings of this study are available on request from the corresponding author. The data are not publicly available due to privacy or ethical restrictions.

**Data sharing statement**

No additional data are available.

**Open access**

This is an open-access article distributed under the terms and conditions of the Creative Commons Attribution license (http://creativecommons.org/licenses/by/4.0/).

**Copyrights**

Copyright for this article is retained by the author(s), with first publication rights granted to the journal.

**References**

Affes, Z., & Hentati-Kaffe, R. (2016). *Predicting US banks bankruptcy: logit versus Canonical Discriminant analysis.* Retrieved from https://ssrn.com/abstract=2826599

Al-Sharkas, A. A., & Al-Sharkas, T. A. (2022). *The impact on bank profitability: Testing for capital adequacy ratio, cost-income ratio and non-performing loans in emerging markets.* Journal of Governance & Regulation, 11(1), 231-243. https://doi.org/10.22495/jgrv11i1siart4

Altman, E. I. (1968). *Financial Ratios, Discriminant Analysis and The Prediction of Corporate Bankruptcy.* The Journal of Finance, 23(4). http://dx.doi.org/10.2307/2978933

Altman, E. I. (2000). *Predicting financial distress of companies: Revisiting the z-score and zeta models.* Retrieved from https://pages.stern.nyu.edu/~ealtman/Z-scores.pdf

Associação Moçambicana de Bancos. (2016). *Banco Central intervém na banca comercial e salvaguarda os direitos dos clientes*: Boletim Informativo da Associação Moçambicana de Bancos, Maputo, 9, 1-16. Retrived from http://www.amb.co.mz/index.php/estudos-e-publicacoes/amb-newsletter/69--27/file.

Ayinla, A. S., & Adekunle, B. K. (2015). *An Overview and Application of Discriminant Analysis in Data Analysis*. IOSR Journal of Mathematics (IOSR-JM), 11(1), 12-15. Retrieved from www.iosrjournals.org

Babajide, A. A., Olokoyo, F. O., & Adegboye, F. B. (2015). *Predicting Bank Failure in Nigeria Using Survival Analysis Approach*. Journal of South African Business Research, Vol. 2015 (2015). http://doi:10.5171/2015.965940

Betz, F., Oprică, S., Peltonen, T., & Sarlin, P. (2013). *Predicting Distress in European Banks: Macroprudential Research Network (European Central Bank) 35.* Retrieved from https://www.ecb.europa.eu/pub/pdf/scpwps/ecbwp1597.pdf

Citterio, A. (2020). *Bank failures: review and comparison of prediction models.* Department of Economics, University of Insubria, Via Monte Generoso 71, Varese 21100, Italy. Retrieved from https://ssrn.com/abstract=3719997

Dwyer, D., Kocagil, A., & Stein, R. (2004). *Moody's KMV Riskcalc v3.1 model. Next generation technology for predicting private firm credit risk.* Retrieved from https://www.moodys.com/sites/products/productattachments/riskcalc%203.1%20whitepaper.pdf

Dwyer, D., Guo, G., & Hood, F. I. (2006). *Moody's KMV Riskcalc v3.1 U.S. Banks. Moody's KMV Company*. Retrieved from https://www.moodys.com/sites/products/ProductAttachments/RiskCalc%20Version%203.1%20U.S.%20Banks.pdf

Central Bank of Mozambique. (2015). *Annual Report*. Retrieved from https://www.bancomoc.mz/media/npopmkmb/_pt_404_rap-2015-pt.pdf

Central Bank of Mozambique. (2020). *Annual Report*. Retrieved from https://www.bancomoc.mz/media/qrzb0rzp/_pt_373_relatorio-anual-2020.pdf

Fejér-Király, G. (2015). Bankruptcy Prediction: A Survey on Evolution Critiques, and Solutions. *ActaUniv. Sapientiae, Economics and Business, 3*(2015), 93-108. http://doi:10.1515/auseb-2015-0006

Gumbo, V., & Zoromedza, S. (2016). Bank Failure Prediction Model for Zimbabwe. *Applied Economics and Finance, 3*(3), 222-235. https://doi.org/10.11114/aef.v3i3.1639. https://doi.org/10.11114/aef.v3i3.1639

Iannotta, G., Nocera, G., & Sironi, A. (2007). Ownership structure, risk and performance in the European banking industry. *Journal of Banking & Finance, 31*(7), 2127-2149. https://doi.org/10.1016/j.jbankfin.2006.07.013

Klynveld Peat Marwick Goerdeler. (2014). *Banking Survey. Financial Services*. Mozambique. Retrieved from https://amb.co.mz/wp-content/uploads/2023/01/Banking-Survey-2014.pdf

Klynveld Peat Marwick Goerdeler. (2015). *Banking Survey. Financial Services*. Mozambique. Retrieved from https://www.acismoz.com/wp-content/uploads/2017/06/Pesquisa%20Bancaria%202015.pdf

Klynveld Peat Marwick Goerdeler. (2016). *Banking Survey. Financial Services*. Mozambique. Retrieved from https://amb.co.mz/wp-content/uploads/2022/10/Banking-Survey-2016_compressed.pdf

Klynveld Peat Marwick Goerdeler. (2019). *Banking Survey. Financial Services*. Mozambique. Retrieved from https://amb.co.mz/wp-content/uploads/2022/12/Banking-Survey-2019.pdf

Klynveld Peat Marwick Goerdeler. (2022). *Banking Survey. Financial Services*. Mozambique. Retrieved from https://amb.co.mz/wp-content/uploads/2023/02/Banking-Survey-2022.pdf

Naili, M., & Lahrichi (2022). *Banks' credit risk, systematic determinants and specific factors: recent evidence from emerging markets*. Heliyon Journal. https://doi.org/10.1016/j.heliyon.2022.e08960

Obadire, A. M., Moyo, V., & Munzhelele, N. F. (2022). Basel III Capital Regulations and Bank Efficiency: Evidence from Selected African Countries. *International Journal of Financial Studies, 10,* 57. https://doi.org/10.3390/ijfs10030057

Peres, C., & Antão, M. (2017). *The use of multivariate discriminant analysis to predict corporate bankruptcy: a review.* AESTIMATIO, the IEB international journal of finance, 14, 108-131. http://doi:10.5605/IEB.14.6

Poulsen, J., & French, A. (2018). *Discriminant Function Analysis*. Retrieved from http://www.statisticssolutions.com/discriminant-analysis

Santillan-Salgado, R. J. (2015). Global Regulatory Changes to the Banking Industry after the Financial Crisis: Basel III. *Journal of Global Economy, 11*(2). http://dx.doi.org/10.2139/ssrn.2692045

Singhal, N., Goyal, S., Sharma, D., Kumari, S., & Nagar, S. (2022). Capitalization and profitability: applicability of capital theories in BRICS banking sector. *Future Business Journal.* https://doi.org/10.1186/s43093-022-00140-w

Stella, O. (2019). *Discriminant Analysis: An Analysis of Its Predictship Function*. Journal of Education and Practice. https://doi:10.7176/JEP

Voz da America. (2016a, November 18). *CIP diz que mais cinco bancos moçambicanos estão em apuros.* Retrieved from https://www.voaportugues.com/a/mocambique-bancos-cip/3602771.html

Voz da America. (2016b, November 21). *Falência de Nosso Banco alerta sistema bancário moçambicano*. Retrieved from https://www.voaportugues.com/a/falencia-nosso-banco-alerta-sistema-bancario-mocambicano/3605726.html

Wang, Y., Dwyer, D., & Zhao, J. (2014). *Riskcalc Banks v4.0 model. Moody's KMV Company*. Retrieved from https://www.moodysanalytics.com/-/media/whitepaper/2016/2016-01-07-RiskCalc-40-US-Banks.pdf

Word Bank. (2020*). Domestic credit to provate sector (% of GDP) – Mozambique, Zimbabwe, Zambia, Tanzania*. Retrieved from https://databank.worldbank.org/reports.aspx?source=2&series=FS.AST.PRVT.GD.ZS&country=MOZ,ZWE,ZMB,TZA

**Appendix**

Appendix A. Banks intervened by the Central Bank of Mozambique

| BANKRUPT BANKS | | | | | | | | | |
|---|---|---|---|---|---|---|---|---|---|
| Moza Banco. S.A | 2020 | 2019 | 2018 | 2017 | 2016 | 2015 | 2014 | 2013 | 2012 |
| Equity to Average Total Asset (EAA) | 17.74% | 19.89% | 28.83% | 26.49% | 0.00% | 8.96% | 11.07% | 11.50% | 22.02% |
| Return on average Equity (ROAE) | 1.85% | -9.19% | -17.91% | -10.68% | 0.00% | 3.60% | 8.87% | 1.66% | -1.47% |
| Return on Average Total Assets (ROAA) | 0.33% | -1.98% | -4.62% | -2.83% | 0.00% | 0.30% | 0.81% | 0.19% | -1.12% |
| Net Interest Income by Average Total Assets (NII) | 5.27% | 5.95% | 5.89% | 7.24% | 0.00% | 2.70% | 4.11% | 5.54% | 5.89% |
| Loans and Advances to Average Total Assets (L&AAA) | 64.78% | 73.11% | 69.30% | 67.08% | 0.00% | 67.75% | 73.94% | 72.13% | 84.75% |
| Bad Debts to Total Loans and Advances (BDTL&A) | 34.56% | 32.92% | 9.45% | 10.18% | 0.00% | 2.00% | 2.22% | 3.28% | 1.54% |
| *z-score* | ▲ 148.65% | ▲ 108.72% | ▼ -26.51% | ▲ 1.47% | n.a | ▼ -8.98% | ▲ 4.93% | ▬ -2.86% | ▼ -26.92% |
| **Nosso Banco, S.A (Formerly Banco Mercantil e de Investimentos, S.A)** | | | | | | | | | |
| Equity to Average Total Asset (EAA) | 0.00% | 0.00% | 0.00% | 0.00% | 0.00% | 11.74% | 15.85% | -8.85% | -11.27% |
| Return on average Equity (ROAE) | 0.00% | 0.00% | 0.00% | 0.00% | 0.00% | 10.31% | 25.59% | 13.31% | 0.19% |
| Return on Average Total Assets (ROAA) | 0.00% | 0.00% | 0.00% | 0.00% | 0.00% | 1.09% | 1.38% | -1.21% | -6.10% |
| Net Interest Income by Average Total Assets (NII) | 0.00% | 0.00% | 0.00% | 0.00% | 0.00% | 7.50% | 6.00% | 0.05% | 0.00% |
| Loans and Advances to Total Average Assets (L&AAA) | 0.00% | 0.00% | 0.00% | 0.00% | 0.00% | 45.15% | 118.92% | 63.66% | 44.82% |
| Bad Debts to Total Loans and Advances (BDTL&A) | 0.00% | 0.00% | 0.00% | 0.00% | 0.00% | 6.00% | 4.00% | 2.16% | 0.26% |
| *z-score* | n.a | n.a | n.a | n.a | n.a | ▲ 48.71% | ▲ 33.16% | ▲ 5.23% | ▼ -35.29% |

## Appendix B. Banks not intervened by the Central Bank of Mozambique

| NON-BANKRUPT BANKS | | | | | | | | | |
|---|---|---|---|---|---|---|---|---|---|
| **Banco Internacional de Moçambique, S.A** | **2020** | **2019** | **2018** | **2017** | **2016** | **2015** | **2014** | **2013** | **2012** |
| Equity to Average Total Asset (EAA) | 20.21% | 22.19% | 21.72% | 19.58% | 17.14% | 17.17% | 17.71% | 18.25% | 18.23% |
| Return on average Equity (ROAE) | 14.92% | 20.60% | 22.29% | 23.32% | 22.40% | 19.21% | 22.69% | 25.19% | 27.20% |
| Return on Average Total Assets (ROAA) | 3.01% | 4.34% | 4.49% | 4.15% | 3.60% | 3.10% | 3.74% | 4.23% | 4.53% |
| Net Interest Income by Average Total Assets (NII) | 3.00% | 8.00% | 9.00% | 9.00% | 7.00% | 5.30% | 5.90% | 6.20% | 7.00% |
| Loans and Advances to Total Average Assets (L&AAA) | 28.25% | 32.51% | 39.84% | 50.59% | 67.32% | 64.52% | 62.54% | 65.21% | 62.46% |
| Bad Debts to Total Loans and Advances (BDTL&A) | 6.31% | 8.42% | 5.45% | 5.55% | 4.00% | 4.00% | 2.50% | 1.80% | 2.10% |
| z-score | ▲ 62.20% | ▲ 97.51% | ▲ 86.27% | ▲ 82.95% | ▲ 59.43% | ▲ 48.63% | ▲ 53.01% | ▲ 55.71% | ▲ 65.47% |
| **Banco Comercial e de Investimentos, S.A** | | | | | | | | | |
| Equity to Average Total Asset (EAA) | 9.86% | 9.75% | 8.14% | 8.76% | 7.33% | 8.07% | 6.53% | 6.37% | 6.56% |
| Return on average Equity (ROAE) | 16.23% | 24.67% | 31.44% | 21.58% | 15.04% | 22.89% | 26.75% | 32.68% | 66.20% |
| Return on Average Total Assets (ROAA) | 1.51% | 2.18% | 2.61% | 1.66% | 1.06% | 1.53% | 1.59% | 1.89% | 2.17% |
| Net Interest Income by Average Total Assets (NII) | 6.00% | 7.00% | 6.40% | 5.50% | 5.00% | 3.70% | 4.20% | 3.60% | 4.50% |
| Loans and Advances to Total Average Assets (L&AAA) | 42.67% | 45.74% | 46.79% | 50.90% | 63.65% | 63.97% | 66.99% | 61.38% | 63.05% |
| Bad Debts to Total Loans and Advances (BDTL&A) | 12.27% | 11.74% | 5.66% | 8.40% | 4.00% | 2.00% | 1.10% | 1.30% | 1.00% |
| z-score | ▲ 89.89% | ▲ 108.14% | ▲ 93.17% | ▲ 78.32% | ▲ 34.54% | ▲ 39.89% | ▲ 43.87% | ▲ 59.66% | ▲ 132.38% |
| **Standard Bank, S.A** | | | | | | | | | |
| Equity to Average Total Asset (EAA) | 21.64% | 22.74% | 21.30% | 19.94% | 17.16% | 17.66% | 17.05% | 15.68% | 16.04% |
| Return on average Equity (ROAE) | 20.44% | 21.48% | 28.99% | 37.21% | 24.35% | 25.87% | 21.41% | 19.23% | 21.80% |
| Return on Average Total Assets (ROAA) | 4.15% | 4.51% | 5.65% | 6.38% | 3.79% | 4.06% | 3.38% | 2.88% | 3.23% |
| Net Interest Income by Average Total Assets (NII) | 13.00% | 5.00% | 9.00% | 11.00% | 7.00% | 5.00% | 5.50% | 5.50% | 6.10% |
| Loans and Advances to Total Average Assets (L&AAA) | 29.62% | 27.48% | 49.37% | 42.73% | 68.45% | 82.16% | 72.50% | 44.77% | 39.55% |
| Bad Debts to Total Loans and Advances (BDTL&A) | 2.62% | 1.47% | 1.61% | 3.01% | 5.00% | 1.00% | 2.00% | 2.60% | 2.80% |
| z-score | ▲ 82.55% | ▲ 62.24% | ▲ 80.84% | ▲ 115.06% | ▲ 68.30% | ▲ 41.85% | ▲ 41.18% | ▲ 51.64% | ▲ 62.94% |
| **Absa Bank Moçambique SA (Formerly Barclays Bank Moçambique, S.A)** | | | | | | | | | |
| Category (ratios) | | | | | | | | | |
| Equity to Average Total Asset (EAA) | 17.23% | 18.77% | 19.73% | 18.71% | 15.87% | 16.86% | 18.23% | 10.15% | 17.12% |
| Return on average Equity (ROAE) | 3.43% | 14.79% | 22.72% | 22.70% | 14.20% | 8.28% | -0.63% | -28.53% | -34.10% |
| Return on Average Total Assets (ROAA) | 0.58% | 2.65% | 4.04% | 3.81% | 2.12% | 1.37% | -0.08% | -3.62% | -4.94% |
| Net Interest Income by Average Total Assets (NII) | 7.75% | 8.32% | 10.00% | 14.00% | 10.00% | 7.00% | 4.80% | 4.40% | 5.50% |
| Loans and Advances to Total Average Assets (L&AAA) | 53.43% | 43.03% | 34.95% | 35.58% | 52.42% | 50.73% | 53.10% | 47.71% | 43.93% |
| Bad Debts to Total Loans and Advances (BDTL&A) | 4.26% | 4.51% | 6.16% | 2.84% | 6.00% | 5.00% | 7.00% | 10.90% | 9.40% |
| z-score | ▲ 20.71% | ▲ 57.95% | ▲ 94.27% | ▲ 87.18% | ▲ 61.93% | ▲ 36.20% | ▲ 16.35% | ▼ -32.20% | ▼ -50.47% |
| **Banco Terra, S.A** | | | | | | | | | |
| Equity to Average Total Asset (EAA) | 0.00% | 0.00% | 35.78% | 40.36% | 44.71% | 55.95% | 44.07% | 29.66% | 25.68% |
| Return on average Equity (ROAE) | 0.00% | 0.00% | -27.62% | 0.37% | 0.77% | 0.43% | -38.93% | -42.15% | -88.00% |
| Return on Average Total Assets (ROAA) | 0.00% | 0.00% | -11.85% | 0.15% | 0.34% | 0.19% | -14.00% | -11.34% | -18.85% |
| Net Interest Income by Average Total Assets (NII) | 0.00% | 0.00% | 10.00% | 12.00% | 5.00% | 6.90% | 7.10% | 5.92% | 5.58% |
| Loans and Advances to Total Average Assets (L&AAA) | 0.00% | 0.00% | 78.74% | 71.38% | 82.83% | 82.63% | 65.84% | 62.72% | 86.23% |
| Bad Debts to Total Loans and Advances (BDTL&A) | 0.00% | 0.00% | 23.43% | 20.77% | 15.00% | 13.00% | 25.30% | 37.55% | 17.65% |
| z-score | n.a | n.a | ▲ 5.26% | ▲ 91.60% | ▲ 43.24% | ▲ 36.83% | ▼ -16.63% | ▲ 40.50% | ▼ -183.57% |
| **Nedbank Moçambique, S.A (formerly Banco Único, S.A)** | | | | | | | | | |
| Category (ratios) | | | | | | | | | |
| Equity to Average Total Asset (EAA) | 12.39% | 15.30% | 14.38% | 13.66% | 12.19% | 13.42% | 11.19% | 11.82% | 22.72% |
| Return on average Equity (ROAE) | -10.13% | -13.11% | 12.85% | 18.50% | 19.93% | 7.32% | 1.84% | -10.24% | -34.90% |
| Return on Average Total Assets (ROAA) | -1.32% | -1.88% | 1.75% | 2.22% | 2.21% | 0.79% | 0.18% | -1.27% | -6.16% |
| Net Interest Income by Average Total Assets (NII) | -1.00% | 2.00% | 7.00% | 7.50% | 6.00% | 6.20% | 5.70% | 4.80% | 1.30% |
| Loans and Advances to Total Average Assets (L&AAA) | 35.14% | 42.48% | 51.68% | 49.09% | 59.40% | 65.04% | 72.49% | 68.17% | 66.14% |
| Bad Debts to Total Loans and Advances (BDTL&A) | 4.12% | 28.27% | 9.00% | 3.63% | 2.00% | 4.00% | 1.90% | 1.80% | 0.20% |
| z-score | ▼ -25.67% | ▲ 84.25% | ▲ 65.57% | ▲ 56.00% | ▲ 42.79% | ▲ 19.20% | ▼ -8.82% | ▼ -39.03% | ▼ -119.88% |
| **FNB Moçambique, S.A** | | | | | | | | | |
| Equity to Average Total Asset (EAA) | 11.86% | 10.10% | 8.93% | 12.43% | 15.13% | 20.77% | 18.61% | 12.82% | 11.67% |
| Return on average Equity (ROAE) | -8.55% | -19.08% | -19.32% | -11.63% | -13.79% | 11.44% | 13.50% | 13.93% | -7.20% |
| Return on Average Total Assets (ROAA) | -0.88% | -1.79% | -2.06% | -1.54% | -2.24% | 2.02% | 1.89% | 1.51% | -1.75% |
| Net Interest Income by Average Total Assets (NII) | -0.79% | -1.70% | 9.00% | 10.00% | 8.00% | 6.40% | 6.70% | 9.90% | 4.80% |
| Loans and Advances to Total Average Assets (L&AAA) | 18.97% | 22.59% | 38.26% | 38.91% | 54.87% | 64.32% | 68.44% | 73.30% | 56.08% |
| Bad Debts to Total Loans and Advances (BDTL&A) | 7.05% | 16.56% | 38.31% | 38.88% | 16.00% | 8.00% | 2.00% | 2.60% | 3.40% |
| z-score | ▲ 1.14% | ▲ 17.32% | ▲ 136.91% | ▲ 159.42% | ▲ 32.02% | ▲ 50.65% | ▲ 25.15% | ▲ 33.41% | ▼ -20.25% |
| **African Banking Cooperation (Moçambique), S.A** | | | | | | | | | |
| Equity to Average Total Asset (EAA) | 11.78% | 15.46% | 18.63% | 20.21% | 19.16% | 14.23% | 6.11% | 9.25% | 12.41% |
| Return on average Equity (ROAE) | -4.69% | -6.59% | 7.80% | 3.44% | 6.91% | -8.91% | -7.35% | 4.46% | 9.80% |
| Return on Average Total Assets (ROAA) | -0.60% | -1.05% | 1.51% | 0.68% | 1.12% | -0.89% | -0.47% | 0.41% | 1.13% |
| Net Interest Income by Average Total Assets (NII) | 5.10% | 5.30% | 8.00% | 10.00% | 8.00% | 5.80% | 4.40% | 5.60% | 6.00% |
| Loans and Advances to Total Average Assets (L&AAA) | 39.46% | 44.58% | 40.30% | 41.33% | 46.54% | 60.57% | 60.57% | 69.37% | 60.33% |
| Bad Debts to Total Loans and Advances (BDTL&A) | 12.36% | 12.36% | 22.64% | 22.04% | 9.00% | 12.00% | 12.10% | 8.40% | 8.40% |
| z-score | ▲ 39.30% | ▲ 31.91% | ▲ 126.42% | ▲ 116.28% | ▲ 55.88% | ▲ 19.05% | ▲ 20.96% | ▲ 29.54% | ▲ 48.09% |
| **Ecobank Moçambique, S.A** | | | | | | | | | |
| Equity to Average Total Asset (EAA) | 28.65% | 29.94% | 21.45% | 9.75% | 5.51% | 14.32% | 34.49% | 14.27% | 0.00% |
| Return on average Equity (ROAE) | -6.12% | -2.55% | 0.24% | -103.90% | -80.57% | -60.29% | -17.01% | -7.63% | 0.00% |
| Return on Average Total Assets (ROAA) | -1.80% | -0.62% | 0.04% | -7.69% | -7.44% | -12.35% | -3.93% | -1.09% | 0.00% |
| Net Interest Income by Average Total Assets (NII) | 7.00% | 5.00% | 4.00% | 2.00% | 5.00% | 6.30% | 14.20% | 18.00% | 0.00% |
| Loans and Advances to Total Average Assets (L&AAA) | 20.43% | 9.99% | 31.67% | 20.14% | 24.43% | 31.19% | 31.06% | 51.62% | 0.00% |
| Bad Debts to Total Loans and Advances (BDTL&A) | 1.91% | 0.68% | 14.87% | 6.07% | 3.00% | 10.00% | 4.00% | 7.00% | 0.00% |
| z-score | ▬ -3.20% | ▲ 1.97% | ▲ 64.21% | ▼ -219.81% | ▼ -178.30% | ▼ -114.62% | ▼ -10.45% | ▲ 31.02% | n.a |
| **First Capital Bank (Mozambique), S.A** | | | | | | | | | |
| Equity to Average Total Asset (EAA) | 27.60% | 38.46% | 38.16% | 51.33% | 40.16% | 36.26% | 34.50% | 24.97% | 0.00% |
| Return on average Equity (ROAE) | 7.53% | 5.18% | 0.76% | -16.36% | 3.95% | -15.10% | -28.28% | -41.82% | 0.00% |
| Return on Average Total Assets (ROAA) | 1.97% | 1.63% | 0.29% | -6.82% | 1.36% | -4.50% | -7.96% | -10.44% | 0.00% |
| Net Interest Income by Average Total Assets (NII) | 4.79% | 6.68% | 9.90% | 8.34% | 8.00% | 5.50% | 4.00% | 2.00% | 0.00% |
| Loans and Advances to Total Average Assets (L&AAA) | 35.89% | 37.96% | 29.59% | 40.52% | 54.06% | 69.69% | 46.81% | 30.57% | 0.00% |
| Bad Debts to Total Loans and Advances (BDTL&A) | 5.13% | 3.58% | 15.22% | 15.64% | 7.00% | 12.00% | 39.00% | 22.00% | 0.00% |
| *z-score* | ▲ 38.29% | ▲ 28.07% | ▲ 82.00% | ▲ 19.48% | ▲ 36.14% | ▼ -9.49% | ▲ 88.80% | ▼ -23.59% | n.a |
| **Société Générale Moçambique, S.A (formerly The Mauritius Commercial Bank S.A)** | | | | | | | | | |
| Equity to Average Total Asset (EAA) | 15.84% | 20.05% | 17.55% | 24.33% | 16.78% | 36.62% | 0.00% | 23.48% | 22.49% |
| Return on average Equity (ROAE) | 1.82% | 0.05% | -5.95% | -35.53% | -52.12% | -21.35% | 0.00% | 11.60% | 11.30% |
| Return on Average Total Assets (ROAA) | 0.27% | 0.01% | -1.11% | -6.59% | -11.83% | -7.82% | 0.00% | 2.58% | 2.49% |
| Net Interest Income by Average Total Assets (NII) | 5.00% | 5.00% | 4.00% | 5.20% | 5.00% | 3.00% | 0.00% | 4.20% | 5.30% |
| Loans and Advances to Total Average Assets (L&AAA) | 47.78% | 58.01% | 58.68% | 40.44% | 46.53% | 34.58% | 0.00% | 51.39% | 47.60% |
| Bad Debts to Total Loans and Advances (BDTL&A) | 21.00% | 31.00% | 11.04% | 23.31% | 32.00% | 41.00% | 0.00% | 6.90% | 6.50% |
| *z-score* | ▲ 91.95% | ▲ 129.69% | ▲ 16.83% | ▲ 8.79% | ▬ -2.24% | ▲ 117.02% | n.a | ▲ 48.17% | ▲ 49.92% |
| **Banco Letshego, S.A** | | | | | | | | | |
| Equity to Average Total Asset (EAA) | 49.93% | 41.53% | 34.58% | 30.61% | 30.94% | 27.20% | 0.00% | 0.00% | 0.00% |
| Return on average Equity (ROAE) | 16.51% | 19.48% | 17.28% | 14.39% | 23.90% | 36.32% | 0.00% | 0.00% | 0.00% |
| Return on Average Total Assets (ROAA) | 7.19% | 6.81% | 5.14% | 4.11% | 6.57% | 9.88% | 0.00% | 0.00% | 0.00% |
| Net Interest Income by Average Total Assets (NII) | 18.00% | 18.10% | 17.00% | 15.00% | 7.00% | 6.00% | 0.00% | 0.00% | 0.00% |
| Loans and Advances to Total Average Assets (L&AAA) | 89.00% | 87.24% | 98.61% | 101.18% | 96.29% | 98.97% | 0.00% | 0.00% | 0.00% |
| Bad Debts to Total Loans and Advances (BDTL&A) | 4.44% | 4.70% | 4.15% | 4.54% | 12.00% | 14.00% | 0.00% | 0.00% | 0.00% |
| *z-score* | ▲ 72.26% | ▲ 80.33% | ▲ 60.87% | ▲ 48.05% | ▲ 93.43% | ▲ 134.50% | n.a | n.a | n.a |
| **Banco Mais** | | | | | | | | | |
| Equity to Average Total Asset (EAA) | 32.37% | 31.78% | 22.80% | 14.36% | 11.75% | 20.27% | 50.88% | 0.00% | 0.00% |
| Return on average Equity (ROAE) | 9.25% | 20.22% | -70.69% | -18.96% | -65.00% | -75.54% | -48.08% | 0.00% | 0.00% |
| Return on Average Total Assets (ROAA) | 2.86% | 5.14% | -11.37% | -2.25% | -7.35% | -15.36% | -24.46% | 0.00% | 0.00% |
| Net Interest Income by Average Total Assets (NII) | 14.00% | 12.00% | 9.00% | 6.00% | 6.00% | 4.90% | 1.20% | 0.00% | 0.00% |
| Loans and Advances to Total Average Assets (L&AAA) | 57.62% | 57.92% | 56.57% | 74.63% | 77.81% | 86.88% | 37.29% | 0.00% | 0.00% |
| Bad Debts to Total Loans and Advances (BDTL&A) | 12.28% | 8.36% | 10.20% | 3.24% | 3.00% | 1.00% | 0.00% | 0.00% | 0.00% |
| *z-score* | ▲ 89.21% | ▲ 95.19% | ▼ -139.82% | ▼ -53.87% | ▼ -168.47% | ▼ -228.39% | ▼ -183.12% | n.a | n.a |
| **Banco Nacional de Investimentos, S.A** | | | | | | | | | |
| Equity to Average Total Asset (EAA) | 44.78% | 51.23% | 55.09% | 54.11% | 46.42% | 45.06% | 63.05% | 89.17% | 103.32% |
| Return on average Equity (ROAE) | 4.09% | 1.90% | 5.60% | 6.39% | 12.59% | 12.01% | 3.73% | 2.01% | 2.66% |
| Return on Average Total Assets (ROAA) | 1.81% | 1.00% | 2.90% | 3.33% | 5.82% | 5.20% | 2.22% | 1.91% | 2.51% |
| Net Interest Income by Average Total Assets (NII) | 5.46% | 6.00% | 7.00% | 11.00% | 9.00% | 8.50% | 7.60% | 8.83% | 7.54% |
| Loans and Advances to Total Average Assets (L&AAA) | 55.85% | 34.97% | 31.38% | 27.24% | 15.87% | 9.30% | 29.16% | 16.09% | 3.96% |
| Bad Debts to Total Loans and Advances (BDTL&A) | 2.50% | 19.20% | 16.26% | 29.44% | 2.00% | 0.00% | 0.00% | 0.58% | 0.00% |
| *z-score* | ▲ 9.20% | ▲ 92.67% | ▲ 95.53% | ▲ 172.27% | ▲ 63.17% | ▲ 52.94% | ▲ 14.99% | ▲ 21.49% | ▲ 23.95% |
| **Mybucks Banking Corporations, S.A (formerly Opportunity Bank S.A)** | | | | | | | | | |
| Equity to Average Total Asset (EAA) | 7.82% | 12.62% | 32.27% | 34.70% | 19.18% | 27.91% | 0.00% | 0.00% | 0.00% |
| Return on average Equity (ROAE) | -22.16% | -17.18% | 43.58% | -43.71% | -63.90% | -12.86% | 0.00% | 0.00% | 0.00% |
| Return on Average Total Assets (ROAA) | -1.81% | -2.78% | 10.14% | -10.50% | -14.83% | -3.59% | 0.00% | 0.00% | 0.00% |
| Net Interest Income by Average Total Assets (NII) | 13.00% | 21.00% | 28.00% | 26.00% | 37.00% | 35.00% | 0.00% | 0.00% | 0.00% |
| Loans and Advances to Total Average Assets (L&AAA) | 61.82% | 70.33% | 118.18% | 53.52% | 77.67% | 73.57% | 0.00% | 0.00% | 0.00% |
| Bad Debts to Total Loans and Advances (BDTL&A) | 21.00% | 16.00% | 2.39% | 7.34% | 4.00% | 2.00% | 0.00% | 0.00% | 0.00% |
| *z-score* | ▲ 47.57% | ▲ 46.85% | ▲ 138.57% | ▼ -51.69% | ▼ -106.97% | ▲ 18.90% | n.a | n.a | n.a |
| **Socremo Banco de Microfinanças, S.A** | | | | | | | | | |
| Equity to Average Total Asset (EAA) | 42.44% | 40.49% | 40.59% | 38.04% | 35.64% | 29.14% | 28.76% | 28.05% | 24.70% |
| Return on average Equity (ROAE) | 0.80% | 12.63% | 14.04% | 7.68% | 16.75% | 22.35% | 20.24% | 18.99% | 15.64% |
| Return on Average Total Assets (ROAA) | 0.32% | 4.81% | 5.32% | 2.81% | 5.37% | 6.05% | 5.37% | 4.86% | 3.57% |
| Net Interest Income by Average Total Assets (NII) | 19.00% | 30.00% | 19.00% | 26.00% | 25.00% | 24.60% | 26.80% | 27.22% | 25.92% |
| Loans and Advances to Total Average Assets (L&AAA) | 56.23% | 65.57% | 68.83% | 74.28% | 67.23% | 66.47% | 73.48% | 77.35% | 71.19% |
| Bad Debts to Total Loans and Advances (BDTL&A) | 6.78% | 2.60% | 2.98% | 3.09% | 7.00% | 6.00% | 0.00% | 6.69% | 5.49% |
| *z-score* | ▲ 50.69% | ▲ 89.46% | ▲ 67.85% | ▲ 62.38% | ▲ 108.08% | ▲ 116.80% | ▲ 83.97% | ▲ 110.79% | ▲ 94.66% |
| **United Bank for Africa Moçambique, S.A** | | | | | | | | | |
| Equity to Average Total Asset (EAA) | 31.58% | 39.63% | 44.36% | 65.95% | 10.71% | 11.13% | 0.95% | 10.83% | 16.30% |
| Return on average Equity (ROAE) | -7.79% | -3.73% | 0.97% | 0.65% | -180.84% | -3.80% | -10.40% | -13.80% | -70.00% |
| Return on Average Total Assets (ROAA) | -2.56% | -1.57% | 0.43% | 0.24% | -14.55% | -41.94% | -9.80% | -1.53% | -7.95% |
| Net Interest Income by Average Total Assets (NII) | 5.10% | 6.50% | 10.00% | 8.00% | 7.00% | 3.70% | 12.10% | 9.00% | 6.00% |
| Loans and Advances to Total Average Assets (L&AAA) | 8.19% | 2.10% | 1.83% | 10.25% | 38.05% | 67.63% | 66.92% | 60.40% | 44.56% |
| Bad Debts to Total Loans and Advances (BDTL&A) | 1.40% | 28.22% | 23.93% | 84.17% | 82.14% | 47.00% | 0.00% | 26.00% | 9.00% |
| *z-score* | ▼ -9.80% | ▲ 134.39% | ▲ 137.56% | ▲ 411.81% | ▼ -39.59% | ▲ 82.85% | ▼ -51.23% | ▲ 81.01% | ▼ -136.33% |

*Notes. n.a = no data available*
*The red mark reflect high risk of bankrupcy; Yellow reflects uncertainty; Green reflects low rissk of bankrupcy*